\documentclass{svjour3}                     
\RequirePackage{graphicx}
\RequirePackage{mathptmx}      
\RequirePackage{flushend}
\RequirePackage[numbers,sort&compress]{natbib}
\RequirePackage[colorlinks,citecolor=blue,urlcolor=blue,linkcolor=blue]{hyperref}
\RequirePackage[T1]{fontenc}
\journalname{Eur. Phys. J. C}
\smartqed  
\newcommand\beq{\begin{eqnarray}}
\newcommand\eeq{\end{eqnarray}}
\newcommand\bq{\begin{equation}}
\newcommand\eq{\end{equation}}
\begin{document}

\title{Right-handed chirality of muon neutrinos in polarized muon decay at rest 
}


\author{W. Sobk\'ow
}


\institute{W. Sobk\'ow \at Institute of Theoretical Physics, University of Wroc\l{}aw,
Pl. M. Born 9, PL-50-204~Wroc{\l}aw, Poland\label{addr1}
              \\
              Tel.: +48 71-3759-408\\
              Fax: +48 71-321-44-54 \\
              \email{wieslaw.sobkow@ift.uni.wroc.pl}           
}

\date{Received: date / Accepted: date}

\maketitle

\begin{abstract} 
In this paper, we analyze  the polarized muon decay at rest (PMDaR) admitting the non-standard $V+A$ interaction  in addition to the standard V-A interaction.  
We show that the angle-energy distribution of Dirac muon neutrinos ($\nu_{\mu}$') is sensitive to the interference terms between the standard V-A coupling of left-chirality $\nu_{\mu}$ and exotic V+A coupling of right-chirality one, which are proportional to the transverse components of $\nu_{\mu}$ 
spin polarization. The interferences do not vanish in a limit of massless  $\nu_{\mu}$ and include the relative phases to test the CP violation (CPV). It  allows to calculate a neutrino flux  for an assumed configuration of detector in  the case of  SM prediction,  and the upper limits on the neutrino flux  for the left-right  $\nu_{\mu}$ mixture.  It is also demonstrated that the effects of 
  neutrino mass and mixing  are tiny corrections and can be omitted. The most important conclusion is that even in the massless $\nu_{\mu}$ limit, there are physical  effects connected with the mixture of left- and right-handed helicity components in a spin $1/2$ quantum state,  when exotic interaction $V+A$  is admitted contrary to the electron observables, where all the possible interferences vanish. 
\end{abstract}
\section{Introduction}
\label{sec:1}
The PMDaR is a suitable process  to probe the space-time structure and to search for the effects coming from the right-handed chirality of Dirac neutrinos as well as  CPV (in terms of the
CPT invariant theorem, the time reversal violation (TRV) implies CPV and vice versa) in the  purely leptonic charged  weak interactions \cite{Jodidio,LorCP}.   This reaction may also be used to study the question of lepton number violation  and of  neutrino nature \cite{Langacker}.  In order to make these tests feasible, the outgoing neutrinos have to be observed. So far all the tests concerning the muon decay   base on the precise measurements of  the electron (positron) observables and of the (anti)neutrino energy spectrum. It is  worthwhile remarking the high-precision measurements of the angle-energy spectrum of positrons and of the parity violation  made recently by TWIST Collaboration \cite{TWIST}. The KARMEN experiment \cite{KARMEN} has measured  the energy distribution of electron 
neutrinos emitted in positive muon decay.  The possibility of measuring the neutrino energy spectrum has first been proposed by Fetscher \cite{Fetscher92}.  
The obtained results indicate the dominant vector-axial (V-A)  \cite{Gell} structure of the charged current weak interaction according to the standard model (SM) prediction \cite{Glashow,Wein,Salam}. This means that only left chiral massless Dirac neutrinos may take 
part in a charged  weak interaction while  parity violation is maximal. As it is  well known,  the neutrino oscillation experiments
indicate the nonzero neutrino mass and provide first evidence for physics beyond the minimal SM. \\
One should clearly stress that  mentioned above observables include mainly the contributions from the squares of coupling constants of the right-chirality neutrinos and  at most from the interferences within exotic couplings, that  are both very tiny. The transverse components of electron (positron) spin  polarization   contain only the interference terms between the standard vector and nonstandard scalar (and tensor) couplings of left-chirality (anti)neutrinos. All the eventual interference terms  between the standard left-chirality neutrino  and exotic right-chirality neutrino  couplings are strongly suppressed  by  a tiny neutrino mass.  
\\In spite of agreement with all the data within errors, the SM   can not be viewed as a ultimate theory.  One of the fundamental aspects, which are not explained in the SM, is an origin of parity violation at current energies. The another crucial problem  is impossibility to explain  the observed baryon asymmetry of Universe \cite{barion} through  a single CP-violating phase of the Cabibbo-Kobayashi-Maskawa quark-mixing matrix (CKM) \cite{Kobayashi}. Presently the CP violation is observed only in the decays of neutral K- and B-mesons 
\cite{CP}. Till now  there is no direct evidence of the CP violation in the leptonic and semileptonic processes. However, the future superbeam and neutrino factory experiments 
 aim at the measurement of  the CP-violating effects in the lepton
sector, where both neutrino and antineutrino oscillation might be observed \cite{Geer}. Moreover, the flavour violation in the neutrino experiments  as well as the composition and origin of the observed dark matter can not be understood in the framework of SM. It is noteworthy that there are also some hints
in experimental data such as; the anomalies observed in short baseline and reactor neutrino experiments and 
the possibility of existence dark radiation. 
 Above phenomena may be related to the right-handed chirality of neutrinos with different masses, \cite{Drewes}. 
The numerous nonstandard schemes including exotic V+A, scalar (S), tensor (T),  pseudoscalar (P) couplings of  interacting  right-chirality neutrinos and  new sources of  CP-breaking phases have appeared. \\
As the present precise tests do not detect the right-chirality neutrinos,   
it seems  meaningful and natural to search for new tools with the linear terms from the exotic couplings obtained in model-independent way. It would allow to compare the predictions of various nonstandard gauge models  with the experiments. Furthermore, existence of interferences  would enable to look for the TRV effects. 
The relevant tools could be  the neutrino observables carrying  information on the components of (anti)neutrino spin polarization. Presently  such tests are still extremely difficult, because they require the observation of final neutrinos, intensive neutrino beam coming from the polarized source and  efficient neutrino polarimeters. However many experimental groups working with the neutrino beams,  polarized muon decay and artificial  (anti)neutrino sources could try to realize the neutrino polarimetry experiments.  
 \\
In this paper, we analyze a scenario with the right-chirality neutrinos produced in the PMDaR by  $V+A$ interaction at low energies. Such interaction may  occur  in the various versions of left-right symmetric models (LRSM) with  $SU(2)_{L}\times SU(2)_{R}\times U(1)$ as the gauge group \cite{LR,Beg}. 
The LRSM emerged first in the framework of grand unified theories (GUT).  They restore the parity symmetry at high energies and give its  violation at low energies as a result of  gauge symmetry breaking.  
There are many theoretical and experimental papers   devoted to  the  $V+A$ weak interaction effects, lepton flavor  and CPV  in the various leptonic and semileptonic processes, e. g.   
\cite{Herczeg,Delphi,Fetscher,Kuno,Czakon}.\\   
The main goal  is to show how the angle-energy distribution of the Dirac $\nu_{\mu}$'s produced in  the PMDaR depends on the interference terms between the standard vector coupling of the left-chirality neutrinos and exotic vector  couplings of the right-chirality ones in the limit of vanishing $\nu_{\mu}$ mass. 
It  allows to estimate  the flux of $\nu_{\mu}$'s  for a detector in the shape of  flat circular ring with a low threshold, both for the SM prediction and the case of neutrino left-right mixture. We show that the angle-energy neutrino distribution may also be used  to probe the CP-violating phases. In addition, we analyze the effects of  neutrino mass and mixing. 
Our study is not made in the framework of concrete version of the left-right symmetric model. \\
This paper is organized as follows: Sec. II contains the basic assumptions as to the production process of $\nu_{\mu}$ beam. In the sec. III,  the results for the energy-angle distribution of $\nu_{\mu}$'s  coming from the PMDaR are presented. In the sec. IV, we present the numerical results concerning the $\nu_{\mu}$ flux  both for the SM and  the case of left-right chirality $\nu_{\mu}$ mixture,  when  $\nu_{\mu}$ beam may be transversely polarized. Finally, we summarize our considerations.  
\\ The  
results are presented in a limit of infinitesimally small mass for all the particles produced in the PMDaR. 
The density operators  \cite{Michel} for the polarized initial muon and for the
polarized outgoing $\nu_{\mu}$ are used. We use the
system of natural units with $\hbar=c=1$, Dirac-Pauli representation of the
$\gamma$-matrices and the $(+, -, -, -)$ metric \cite{Greiner}. 

\section{Polarized muon decay at rest - muon neutrino beam}
\label{sec:2}
\begin{figure}
  \includegraphics[width=0.8\textwidth]{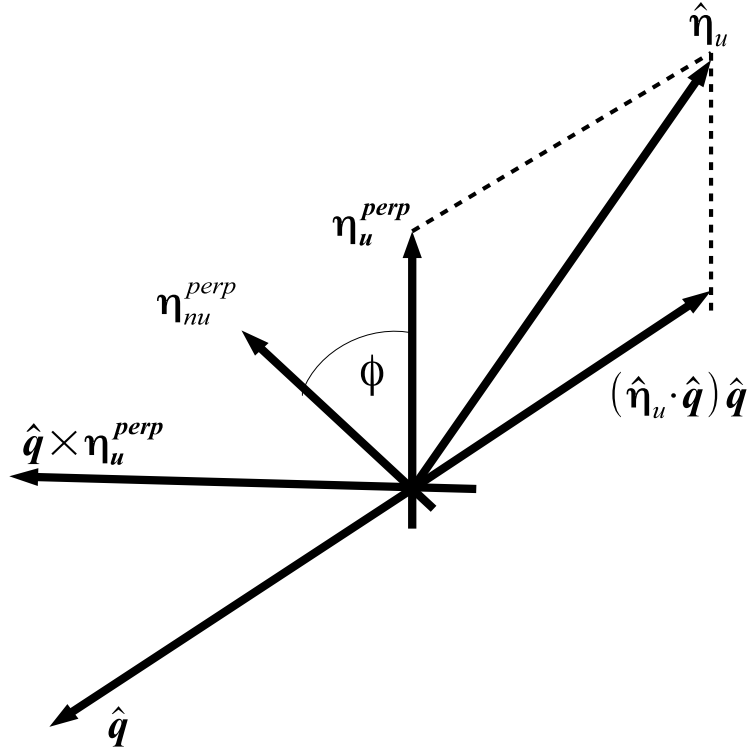}
  \caption{Production plane for $\nu_{\mu}$ produced in PMDaR}
       \label{fig1}
\end{figure}

We  assume that the PMDaR  $(\mu^{-} \rightarrow e^- +
\overline{\nu}_{e} + \nu_{\mu})$ is a source of the Dirac $\nu_{\mu}$ beam. This process is described at a level of lepton-number-conserving four-fermion point interaction. 
We admit a presence of the exotic vector $g_{LR, RL, RR}^V$ 
couplings in addition to the standard vector $g_{LL}^V$ coupling. The indexes $(L, R)$ describe the chirality of the final electron and initial stopped muon, respectively. 
 It means that
the outgoing $\nu_{\mu}$ flux is a mixture of the left-chirality
and  right-chirality $\nu_{\mu}$'s. 
 The  amplitude for the above process is of the form:
\beq
M_{\mu^{-}} & = &
\frac{G_{F}}{\sqrt{2}}\{g_{LL}^{V}(\overline{u}_{e}\gamma_{\alpha}(1-\gamma_5)v_{\nu_{e}})
(\overline{u}_{\nu_{\mu}} \gamma^{\alpha}(1 -
\gamma_{5})u_{\mu}) \nonumber\\
&&  \mbox{} + g_{RR}^{V}(\overline{u}_{e}\gamma_{\alpha}(1+\gamma_5)v_{\nu_{e}})
(\overline{u}_{\nu_{\mu}} \gamma^{\alpha}(1 +
\gamma_{5})u_{\mu})\\
&&  \mbox{} + g_{LR}^{V}(\overline{u}_{e}\gamma_{\alpha}(1-\gamma_5)v_{\nu_{e}})
(\overline{u}_{\nu_{\mu}} \gamma^{\alpha}(1 +
\gamma_{5})u_{\mu})
\nonumber\\
&&  \mbox{} + g_{RL}^{V}(\overline{u}_{e}\gamma_{\alpha}(1+\gamma_5)v_{\nu_{e}})
(\overline{u}_{\nu_{\mu}} \gamma^{\alpha}(1 -
\gamma_{5})u_{\mu}),\nonumber
\}\eeq
where $ v_{\nu_{e}}$ and
$\overline{u}_{e}$ $(u_{\mu}\;$ and $\; \overline{u}_{\nu_{\mu}})$ are the Dirac
bispinors of the outgoing electron antineutrino and electron (initial muon and
final $\nu_{\mu}$), respectively.\\ $G_{F} = 1.1663788(7)\times
10^{-5}\,\mbox{GeV}^{-2} (0.6 ppm)$ (MuLan Collab.) \cite{Mulan} is the Fermi constant. 
 Our analysis is carried out in the limit of massless neutrino, then 
left-chirality $\nu_{\mu}$ posses negative helicity, while the right-chirality one
has positive helicity, see \cite{Fetscher}.  In the
SM, only $g_{LL}^{V}$ is nonzero value. 
The table \ref{table1} displays the current limits for the $g^{V}_{LL}$, $g_{LR, RL, RR}^V$ couplings, \cite{Data}. 
\begin{table}
\caption{ Current limits on the nonstandard couplings}
\label{table1}
\begin{tabular}{lll}
\hline\noalign{\smallskip}
  Coupling constants & SM & Current limits \\
  \noalign{\smallskip}\hline\noalign{\smallskip}
     $|g_{LL}^V|$ & $1$ &   $>0.960$ \\
    $ |g_{LR}^V|$ & $0$ & $<0.025$ \\
    $|g_{RL}^V|$ & $0$ & $<0.104$ \\
    $|g_{RR}^V|$ & $0$ & $<0.031$ \\
    \noalign{\smallskip}\hline
\end{tabular}
\end{table} 
Because we allow for the nonconservation of the combined symmetry CP, all the
coupling constants $g_{LL}^V, g_{LR, RL, RR}^V$   are complex.\\
The initial muon is at rest and polarized. Its polarization is denoted by the unit vector  $\mbox{\boldmath $\hat{\eta}_{\mu}$}$.  
The production plane is spanned by 
$\mbox{\boldmath $\hat{\eta}_{\mu}$}$ and  the outgoing $\nu_{\mu}$  
LAB momentum unit vector ${\bf \hat{q}}$, see Fig. \ref{fig1}.  
%
As it is well known, in 
this plane, the polarization vector $\mbox{\boldmath $\hat{\eta}_{\mu}$}$  can
be expressed, with respect to the $\hat{\bf q}$, as a sum of the longitudinal component of the muon polarization
$(\mbox{\boldmath$\hat{\eta}_{\mu}$}\cdot\hat{\bf q}){\bf \hat{q}} $
 and transverse component of the muon polarization 
$\mbox{\boldmath $\eta_{\mu} ^{\perp}$} $, which is defined as
$ \mbox{\boldmath $\eta_{\mu}^{\perp}$} =
\mbox{\boldmath$\hat{\eta}_{\mu}$}-
(\mbox{\boldmath$\hat{\eta}_{\mu}$}\cdot\hat{\bf q}) {\bf \hat{q}}$.
\\
Calculations are carried out with  use of the covariant 
density matrices   for the polarized initial muon and  the
polarized outgoing $\nu_{\mu}$, \cite{Michel}. 
The formula for the  Lorentz boosted spin polarization 4-vector of  massive $\nu_{\mu}$ 
$S^\prime$ (in the laboratory frame) is as follows:
\beq
S^\prime & = & (S^{\prime 0}, {\bf S^\prime}),\\
S^{\prime 0} & = & \frac{{|\bf q|}}{m_{\nu}}(\mbox{\boldmath
$\hat{\eta}_{\nu}$}\cdot{\bf \hat{q}}),
\\
{\bf S^\prime} & = &
\left(\frac{E_{\nu}}{m_{\nu}}(\mbox{\boldmath
$\hat{\eta}_{\nu}$}\cdot{\bf \hat{q}}){\bf \hat{q}} +
\mbox{\boldmath $\hat{\eta}_{\nu}$} - (\mbox{\boldmath
$\hat{\eta}_{\nu}$}\cdot{\bf \hat{q}}){\bf \hat{q}}\right),
\eeq
where $\mbox{\boldmath $\hat{\eta}_{\nu}$}$ is the unit 3-vector of
the $\nu_{\mu}$ polarization in its rest frame; 
$(\mbox{\boldmath$\hat{\eta}_{\overline{\nu}}$}\cdot\hat{\bf q}){\bf\hat{q}}$ is the longitudinal component of $\nu_{\mu}$ polarization; 
$\mbox{\boldmath$\hat{\eta}_{\nu}$} - (\mbox{\boldmath
$\hat{\eta}_{\nu}$}\cdot{\bf \hat{q}}){\bf \hat{q}}
= \mbox{\boldmath$\eta_{\nu}^{\perp}$} $ is the transverse component of $\nu_{\mu}$ polarization. 
The formula for the covariant density
matrix of the polarized $\nu_{\mu}$ in the massless limit   is given by:
\beq
\lim_{m_{\nu}\rightarrow 0}\bigg\{\Lambda_{\nu}^{(s)}\equiv u({\bf q}, \mbox{\boldmath $\hat{\eta}_{\nu}$})\overline{u}({\bf q}, \mbox{\boldmath $\hat{\eta}_{\nu}$})\bigg\} &=&
\lim_{m_{\nu}\rightarrow 0}
\frac{1}{2}\bigg\{\left[(q^{\mu}\gamma_{\mu}) + m_{\nu}\right]\left[1 +
\gamma_{5}(S^{\prime \mu}\gamma_{\mu})\right]\bigg\}\nonumber  \\
  =    \frac{1}{2}&\bigg\{&(q^{\mu}\gamma_{\mu})
 \left[1 + \gamma_{5}(\mbox{\boldmath
$\hat{\eta}_{\nu}$}\cdot{\bf \hat{q}}) + \gamma_{5}
S^{\prime\perp}\cdot \gamma \right]\bigg\},
\eeq
where $u({\bf q}, \mbox{\boldmath $\hat{\eta}_{\nu}$})$ is the positive energy plane wave massive spinor solution to Dirac's equation of spin polarization vector $\mbox{\boldmath $\hat{\eta}_{\nu}$}$; $S^{\prime\perp} =
(0, \mbox{\boldmath $\eta_{\nu}^{\perp}$})$. We see
that in spite of the singularity $m_{\nu}^{-1}$ in 
$S^\prime $, the density matrix  $\Lambda_{\nu}^{(s)}$ 
including $\mbox{\boldmath $\eta_{\nu}^{\perp}$}$ remains finite. 
It is worthwhile pointing out that the last term in the projector has different $\gamma$-matrix structure from that of the longitudinal polarization contribution. This term will be responsible for the nonvanishing interference between the standard coupling of left-chirality  $\nu_{\mu}$'s and exotic coupling of right-chirality ones in the angle-energy distribution of $\nu_{\mu}$'s from the PMDaR in the massless limit. 

\section{Energy-angle distribution of muon neutrinos}
\label{sec:3}
In this section, we show how the energy-angle distribution of  $\nu_{\mu}$'s  depends  
on the interference terms between the standard   and exotic
 couplings in the limit of vanishing $\nu_{\mu}$ mass. \\
The  proper formula, obtained  after the integration over all the momentum directions of the outgoing electron and electron antineutrino,  is of the form:
\bq
   \label{didera}
\frac{d^2 \Gamma}{ dy d\Omega_\nu}
  =  
\left(\frac{d^2 \Gamma}{dy
d\Omega_\nu}\right)_{(LL + LR + RL + RR)} 
 + \left(\frac{d^2 \Gamma}{dy
d\Omega_\nu}\right)_{(I)}
\eq 
\beq
\label{kwa}
\left(\frac{d^2 \Gamma}{ dy d\Omega_\nu}\right)_{(LL + LR + RL + RR)}   =  
\frac{G_{F}^2 m_{\mu}^5 }{768\pi^4} \bigg[ (1-\mbox{\boldmath
$\hat{\eta}_{\nu}$}\cdot\hat{\bf q}) 
(|g_{LL}^{V}|^2+|g_{RL}^{V}|^2) && \nonumber\\
 \cdot y^2( -2y+3 -(2y-1)\mbox{\boldmath $\hat{\eta}_{\mu}$}\cdot\hat{\bf q})  + (1+\mbox{\boldmath
$\hat{\eta}_{\nu}$}\cdot\hat{\bf q})(|g_{LR}^{V}|^2 +|g_{RR}^{V}|^2) && \\
\cdot y^2( -2y+3 + (2y-1)\mbox{\boldmath
$\hat{\eta}_{\mu}$}\cdot\hat{\bf q})
 \bigg],&& \nonumber\\
  \left(\frac{d^2 \Gamma}{dy d\Omega_\nu}\right)_{(I)} = \label{INT}
\frac{G_{F}^2 m_{\mu}^5}{384\pi^4} y^2  
  \bigg[\left( Re(g_{LL}^V g_{LR}^{V*}) + Re(g_{RL}^V g_{RR}^{V*})\right)(\mbox{\boldmath
$\eta_{\nu}^{\perp}$}\cdot \mbox{\boldmath
$\hat{\eta}_{\mu}$}) &&  \nonumber \\
 - \left(Im(g_{LL}^V g_{LR}^{V*})  +  Im(g_{RL}^V g_{RR}^{V*})\right)  \mbox{\boldmath $\eta_{\nu}^{\perp}$}\cdot({\bf \hat{q}} \times
\mbox{\boldmath $\hat{\eta}_{\mu}$})  \bigg].&& 
 \eeq
Here, $y=2E_\nu/m_{\mu}$ is the reduced $\nu_\mu$ energy
for the muon mass $m_\mu$, it varies from $0 $ to $1$, and
$d\Omega_\nu$ is the solid angle differential for $\nu_\mu$
momentum $\hat{\bf q}$.
\\
Equation (\ref{INT}) includes the interference term between the 
$g_{LL}^{V}$ (left-chirality $\nu_\mu$) and exotic $g_{LR}^V$ (right-chirality $\nu_\mu$) couplings, so it is linear in the
exotic coupling contrary to Eq. (\ref{kwa}) and  the interference between the exotic  $g_{RL}^V$ and  $g_{RR}^V$ couplings.   
We  see that 
the interference terms do not vanish in the massless $\nu_\mu$ limit. 
It is necessary to stress that there is the different dependence on the $y$ between  quadratic terms and interferences.  
For  $\mbox{\boldmath $\hat{\eta}_{\mu}$} \cdot {\bf \hat{ q}}= 0$, the
interference part can be rewritten in the following way:
 \beq\label{DDR}
 \left(\frac{d^2 \Gamma}{dy d\Omega_\nu}\right)_{(I)} & = &
\frac{G_{F}^2 m_{\mu}^5 }{384\pi^4} |\mbox{\boldmath
$\eta_{\nu}^{\perp}$}| |\mbox{\boldmath $\eta_{\mu} ^{\perp}$}|
  \mbox{} \bigg\{ |g_{LL}^V ||g_{LR}^{V}|cos(\phi + \alpha)
+ |g_{RL}^{V}||g_{RR}^{V}|cos(\phi + \beta)\bigg\}y^2,  
\eeq
where  $\phi$ is the angle between  $\mbox{\boldmath
$\eta_{\nu}^{\perp}$}$ and $\mbox{\boldmath$\eta_{\mu} ^{\perp}$}$ only, Fig. \ref{fig1};  
$\alpha \equiv
\alpha_{V}^{LL} - \alpha_{V}^{LR}$, $\beta \equiv
 \alpha_{V}^{RL} -
\alpha_{V}^{RR} $ are the relative phases between the $g^{V}_{LL}$,  $g^{V}_{LR}$, and   $ g^{V}_{RL}, g^{V}_{RR}$ couplings, respectively. \\
It can be noticed that the relative phases
$\alpha, \beta$ different from $0, \pi$ would indicate the CP
violation in the CC weak interaction.  We see that even in the massless limit the helicity structure of interaction vertices may allow for a helicty flip provided the quantity $\mbox{\boldmath $\eta_{\nu}^{\perp}$}$, which is left invariant under Lorentz boost.  The interference part, Eq.
(\ref{DDR}), includes only the contributions from the transverse component of
the initial muon polarization $\mbox{\boldmath $\eta_{\mu} ^{\perp}$} $ and the
transverse component of the outgoing neutrino polarization $\mbox{\boldmath
$\eta_{\nu}^{\perp}$}$. Both transverse components are perpendicular
with respect to the $\hat{\bf q}$.
\\
Using the current data \cite{Data}, we calculate the upper limit on the
magnitude of $\mbox{\boldmath$\eta_{\nu}^{\perp}$} $ and lower bound for the
$(\mbox{\boldmath$\hat{\eta}_{\nu}$}\cdot\hat{\bf q})$, see \cite{Fetscher}:
\beq \label{trlo}
 |\mbox{\boldmath $\eta_{\nu}^{\perp}$}|
&=& 2\sqrt{Q_{L}^{\nu}(1-Q_{L}^{\nu})}  \leq 0.082, \\
|\mbox{\boldmath $\hat{\eta}_{\nu}$}\cdot\hat{\bf q}| &=& |1- 2
Q_{L}^{\nu}|  \geq 0.997,
\\
Q_{L}^{\nu} &=& 1 - |g_{RR}^V|^{2} -  |g_{LR}^V|^{2}\geq
0.998,
\eeq
where $Q_{L}^{\nu}$ is the probability of finding  the  left-chirality $\nu_{\mu}$.
\\
If the neutrino beam comes from the unpolarized muon decay at rest, there is no interference terms in the $\nu_\mu$ spectral function. \\
It is noteworthy  that the effects coming from the
  neutrino mass and mixing are very tiny and they may be neglected. In order to show this, we use the final
density matrix for the mass eigenstates $m_1, m_2$ of $\nu_{\mu}$ to avoid
breaking the fundamental principles of Quantum Field Theory. We assume that at 
the neutrino detector (target)
  $\nu_\mu = cos \theta \nu_1 + sin \theta
\nu_2$. In this way, the differential $\nu_{\mu}$ spectrum is of the
form: \beq \label{mix} \frac{d^2 \Gamma}{ dy d\Omega_\nu} &=& cos^2 \theta
\frac{d^2 \Gamma}{ dy_1 d\Omega_\nu} + sin^2 \theta \frac{d^2 \Gamma}{dy_2
d\Omega_\nu}= \gamma_{(V)}^{(\phi)}\bigg[ y_{1}^2 + sin^{2} \theta \frac{\delta m_{\nu}^{2}}{m_{\mu}^2} +
O(\frac{\delta m_{\nu}^{2}}{m_{\mu}^2})\bigg], \eeq where  $\gamma_{V}^{(\phi)}= \frac{G_{F}^2 m_{\mu}^5 }{384\pi^4}
    |\mbox{\boldmath $\eta_{\nu}^{\perp}$}|
|\mbox{\boldmath $\eta_{\mu} ^{\perp}$}| \bigg\{ |g_{LL}^V ||g_{LR}^{V}|cos(\phi + \alpha)$
$+|g_{RL}^{V}||g_{RR}^{V}|cos(\phi + \beta)\bigg\}$. Taking into account the available data,  
 the linear contribution from the mass mixing $\frac{\delta m_{\nu}^{2}}{m_{\mu}^2}$ is viewed as extremely small $(\sim 10^{-19})$.   
\\It is worthwhile remarking that if one admits the production of  $\nu_\mu$'s with right-handed chirality by the exotic scalar and  tensor  interactions, there is no interferences between the standard vector and exotic couplings in the limit of vanishing $\nu_\mu$ mass. When one computes the  angle-energy distribution of electron antineutrinos, the mentioned interference terms appear and do not vanish in the massless limit (analysis made in \cite{sobkow}). 
\section{Neutrino flux}
\label{sec:4}
\begin{figure}
\begin{center}
\includegraphics[scale=.8]{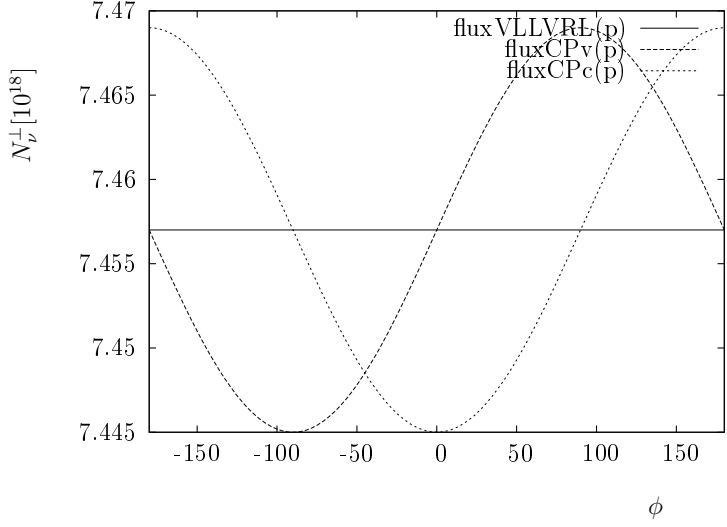}
\end{center}
\caption{Plot of  $N_{\nu}^{\perp} $ as a function  of  $\phi$: CP violation, $\alpha=\beta=3\pi/2$ (long-dashed line);  CP  conservation $\alpha=\beta=\pi$ (short-dashed line);  solid line concerns contribution from  $g_{LL}^{V}, g_{RL}^{V}$  couplings (left-chirality $\nu_{\mu}$).} 
\label{Fig2}
\end{figure}

In order to find the $\nu_{\mu}$ flux, we assume that the hypothetical  detector has  the shape of flat circular ring with a low threshold  $T_{e}^{th}=10 \,eV$ $(y_{e}^{th}=0.0062)$ corresponding to  $E_{\nu}^{min}=1603.44  \, eV$. In addition, one assumes that neutrino source is located in the center of the ring detector and  polarized perpendiculary to the ring. It means that  we must integrate the angle-energy distribution of $\nu_\mu$'s over the neutrino energy in the range $[1603.44  \, eV, m_{\mu}/2]$, and over  
$d\Omega_\nu$ in the proper range, i.e. $\phi_{\nu} \in [0,2\pi]; \theta_{\nu} \in [\pi/2 -\delta, \pi/2 + \delta]$. Then, it is  multiplied  by  $N_{\mu}/S_{D}$ yet, where $N_{\mu}=10^{21}$ is the number of polarized muons decaying per one year.  $S_{D}=4 \pi R^2 sin \delta$, where $R=L=2205 \,cm$ is the inner radius of the detector that is equal to the distance $L$ between  the muon neutrino source and detector, $\delta = 0.01$. In this way one gets the information on  the number of  $\nu_\mu$'s passing through $S_{D}$ in the direction perpendicular to  $\mbox{\boldmath $\hat{\eta}_{\mu}$}$. 
 Fig. \ref{Fig2} displays the dependence of neutrino flux $N_{\nu}^{\perp}$ on the $\phi$ angle for the case of CP conservation (short-dashed line) and of CP violation (long-dashed line) in the presence of $V+A$ interaction, when  
$\hat{\bf q }\perp \mbox{\boldmath $\hat{\eta}_{\mu}$}$. The  solid line describes only  the contribution from the left chiral  $\nu_{\mu}$, i. e. from  $g_{LL}^{V}, g_{RL}^{V}$  couplings.  
\par Using the available data \cite{Data}, we get the flux of $\nu_{\mu}$ beam for the SM (per one year): 
$ \Phi_{\overline{\nu}}^{\perp (SM)}=7.48 \cdot 10^{18}cm^{-2}year^{-1}$, when $\mbox{\boldmath $\hat{\eta}_{\nu}$}\cdot\hat{\bf q} =-1, g_{LL}^{V}=1$.  The lower limit on the  flux of left-chirality $\nu_{\mu}$'s in the presence of exotic couplings is 
$ \Phi_{\overline{\nu}}^{\perp (g_{LL}^{V},g_{RL}^{V})}=7.457 \cdot 10^{18}cm^{-2}year^{-1}$. 
The upper bound on the magnitude  of neutrino flux connected with the interferences $g_{LL}^{V}g_{LR}^{V}$, \\ $g_{RL}^{V}g_{RR}^{V}$ is  $\Phi_{\overline{\nu}}^{\perp,inter}= 1.2\cdot 10^{16}cm^{-2}year^{-1}$, when CP symmetry is violated, i. e. $\phi=\pi/2, \alpha=\beta=3\pi/2$. The same value can be obtained for CP conservation assuming for example $\phi=\pi, \alpha=\beta=\pi$. 
We see that the estimated interference effect is two orders smaller compared to value for the  left chiral  $\nu_{\mu}$'s. 
The numerics are made with the normalized  couplings. 
\section{Conclusions}
\label{sec:5}
 In this paper, we have investigated the PMDaR  in the presence of the exotic $V+A$ interaction, when the (anti)neutrinos have Dirac nature and may have right chirality.\\ We have shown that the angle-energy distribution of $\nu_{\mu}$'s produced in the PMDaR includes the terms with interference between the standard $g_{LL}^{V}$ (left-handed chirality of $\nu_\mu$) and exotic $g_{LR}^V$ (right-handed chirality of $\nu_\mu$) couplings, which are independent of the $\nu_{\mu}$ mass.  These interferences are exclusively proportional to the T-even and T-odd  transverse components of $\nu_{\mu}$ polarization $\mbox{\boldmath $\eta_{\mu} ^{\perp}$}$. 
It is necessary to point out that  if the interacting left- and right-chirality $\nu_{\mu}$ are produced in  the PMDaR, the $\nu_{\mu}$ polarization vector may acquire the transversal component. \\
Next, we have computed the flux of $\nu_{\mu}$ beam for the assumed configuration of detector in  the case of  SM prediction,  and the upper limits on the $\nu_{\mu}$  flux in  the case of left-right $\nu_{\mu}$  mixture. \\ 
We have also displayed that the eventual effects connected with the neutrino mass and mixing in the angle-energy distribution of $\nu_{\mu}$'s are  inessential $( \sim 10^{-19})$, when the distance  between  the $\nu_{\mu}$ source and detector is relatively small $L\sim 22 m$.   \\
The measurements of neutrino observables  seem to be a real challenge for experimental groups, but could shed some more light on  the existence of  right chiral interacting (anti)neutrinos and of the  nonstandard CP-violating phases in the leptonic processes than the current experiments with the electron (positron) observables. 
Let us remind that in the electron observables,  the  eventual interference terms between the standard and exotic $V+A$ couplings are strongly suppressed by the tiny (anty)neutrino mass. One should remark that the appearance of non-vanishing interference term in the massless neutrino limit between the non-standard scalar  $g_{RR}^{S}$ and   standard $g_{LL}^{V}$ couplings in the transversal electron polarization is possible. However in this case, the neutrinos have only the left-handed chirality.  \\
It is worthwhile emphasizing that  there is a well known technique of producing the  polarized muons at rest  (TWIST, PSI, KARMEN, BooNE  Collaborations). A great experience of mentioned  laboratories should be used  in studies on the  efficient neutrino polarimeter. 
 The PMDaR is not a unique weak process, where the right chiral interacting neutrinos can be generated, so it is meaningful to look   for the other polarized (anti)neutrino sources.\\ 
Finally, one ought to point out that even in the massless $\nu_{\mu}$ limit,  there are physical and in principle observable effects coming from the mixture between the left- and right-handed helicity components in the spin $1/2$ quantum state,  when the exotic $V+A$ interaction   is admitted.




\end{document}